# Indication of Collective Flow and Transparency in p-p Collisions at LHC

**Inam–ul Bashir[1]**
*Department of Physics*
*Jamia Millia Islamia*
*New Delhi – 110025 (INDIA)*
E-mail: `inamhep@gmail.com`

**Saeed Uddin**
*Department of Physics*
*Jamia Millia Islamia*
*New Delhi – 110025 (INDIA)*
E-mail: `saeed_jmi@yahoo.co.in`

**Abstract** . The mid-rapidity transverse momentum spectra of various hadrons and the available rapidity distributions of some strange hadrons produced in p-p collisions at LHC energy $\sqrt{s_{NN}}$ = 7.0 TeV have been studied using a Unified Statistical Thermal Freeze-out Model (USTFM). The calculated results are found to be in good agreement with the experimental data. The theoretical fits of the transverse momentum spectra using the model calculations provide the thermal freeze-out conditions in terms of the temperature and collective flow parameters for different hadronic species. The study reveal the presence of significant collective flow and a well defined temperature in the system thus indicating the formation of a thermally equilibrated hydrodynamic system in p-p collisions at LHC. Moreover, the fits to the available experimental rapidity distributions data of strange hadrons show the effect of almost complete transparency in p-p collisions at LHC. The transverse momentum distributions of protons and kaons produced in p-p collisions at $\sqrt{s_{NN}}$ = 200 GeV and $\sqrt{s_{NN}}$ = 2.76 TeV have also been reproduced successfully. The model incorporates longitudinal as well as a transverse hydrodynamic flow. The contributions from heavier decay resonances have also been taken into account. We have also imposed the criteria of exact strangeness conservation in the system.

---

[1]Speaker

**1.Introduction**

Ultra-relativistic heavy-ion collisions at the Large Hadron Collider (LHC) produce strongly interacting matter under the extreme conditions of temperature and energy density, similar to the conditions prevailing during the first few microseconds of the Universe after the Big Bang [1]. The study of multi particle production in ultra-relativistic heavy-ion collisions also allow us to learn the final state distribution of baryon numbers at the thermo-chemical freeze-out after the collision – initially carried by nucleons only before the nuclear collision [2]. Within the framework of the statistical model it is assumed that initially a fireball, i.e. a hot and dense matter of the partons (quarks and gluons) is formed over an extended region after the collision. The quarks and gluons in the fireball may be nearly free (deconfined) due to the ultra-violet freedom i.e. in a quark gluon plasma (QGP) phase. This fireball undergoes a collective expansion accompanied by further particle production processes through the secondary collisions of quarks and gluons which consequently leads to a decrease in its temperature. Eventually the expansion reaches a point where quarks and gluons start interacting non-perturbatively leading to the confinement of quarks and gluons through the formation of hadrons, i.e. the so called hadronization process. In this hot matter which is in the form of a gas of hadronic resonances at high temperature and density, the hadrons continue to interact thereby producing more hadrons and the bulk matter expands further due to a collective hydrodynamic flow developed in the system. This consequently results in a further drop in the thermal temperature because a certain fraction of the available thermal energy is converted into directed (collective hydrodynamic) flow energy. As the mean free paths for different hadrons, due to expansion increases, the process of decoupling of the hadrons from the rest of the system takes place and the hadron spectra are frozen in time. The hadrons with smaller cross-sections stop interacting with the surrounding matter earlier and hence decouple earlier. Hence a so called *sequential* thermal/kinetic freeze-out of different hadronic species occurs. Following this, the hadrons freely stream out to the detectors. The freeze-out conditions of a given hadronic specie are thus directly reflected in its transverse momentum and rapidity spectra [3]

Within the framework of the statistical model [4] the measured particle ratios can be used to ascertain the system temperature and the baryonic chemical potential at the final freeze-out i.e. at the end of the evolution of the hadronic gas phase. The statistical model thus assumes that the system is in thermal and chemical equilibrium at this stage. The system at freeze-out can be described in terms of a nearly free gas of various hadronic resonances (HRG). The above assumptions are valid with or without the formation of a QGP at the initial stage. It is believed that the produced hadrons also carry information about the collision dynamics and the subsequent space-time evolution of the system. Hence precise measurements of the transverse momentum distributions of identified hadrons along with the rapidity spectra are essential for the understanding of the dynamics and properties of the created matter up to the final freeze-out [5]. The transverse momentum distributions are believed to be encoded with the information about the collective transverse and longitudinal expansions and the thermal temperature at freeze-out.

The particle production in p-p collisions are very important as these can serve as a baseline for understanding the particle production mechanism and extraction of the signals of QGP formation in heavy ion collisions [12]. The value of chemical potential is always lower in p-p collisions than in heavy ion collisions due to the lower stopping power in p-p collisions [13]. As one goes to LHC energies, the stopping reduces much further giving rise to nearly zero net baryon density at mid-rapidity and thus the value of the chemical potential at mid-rapidity essentially reduces to zero. Thus at LHC, we believe the p-p collisions to be completely transparent.

The p-p collisions at lower energies were successfully described in the past by using statistical hadronization model [14, 15]. The same kind of analysis has been performed on the p-p results at LHC energy of $\sqrt{s_{NN}}$ = 0.9 TeV [16]. Naively, the p-p collisions are not expected to form QGP or a system with collective hydrodynamic effects. An absence of radial flow in p-p collisions at $\sqrt{s_{NN}}$ = 200 GeV and $\sqrt{s_{NN}}$ = 540 GeV was found in a recent work [17]. However, there have been speculations [18–21] about the possibility of the formation of such a system but of smaller size in the p-p collisions. The occurrence of the high energy density events in high multiplicity p - $\bar{p}$ collisions [22, 23] at CERN-SPS motivated searches for hadronic deconfinement in these collisions at $\sqrt{s_{NN}}$ = 0.54 TeV at SPS [18] and at $\sqrt{s_{NN}}$ = 1.8 TeV [19, 20] at the Tevatron, Fermilab. A common radial flow velocity for meson and anti-baryon found from the analysis of the transverse momentum data of the Tevatron [19] had been attributed to as an evidence for collectivity due to the formation of QGP [24].

Keeping in view the above facts, we in our present analysis will address the collective effect signatures in the p-p collisions at LHC, particularly in terms of transverse and longitudinal flows while attempting to reproduce the transverse momentum and rapidity distributions of hadrons produced in p-p collisions at LHC.

## 2. Model

Though the details of the model used here can be found elsewhere [5-11] however for the sake of convenience we will briefly describe our model here. In a statistical hydrodynamic description taking into account the flow in the transverse and longitudinal directions, the final state particles will leave the hadronic resonance gas (HRG) at the time of freeze-out. The momentum distributions of hadrons, emitted from within an expanding fireball, assumed to be in the state of local thermal equilibrium, are characterized by the Lorentz-invariant Cooper-Frye formula [25]

$$E\frac{d^3N}{d^3p} = \frac{g}{(2\pi)^3}\int f\left(\frac{p^\mu u^\mu}{T}, \lambda\right) p^\mu d\Sigma_\mu \qquad (1)$$

Where g is the degeneracy factor of a given hadronic specie in the expanding relativistic hadronic gas. In recent works [26,27] it has been clearly shown that there is a strong evidence of increasing baryon chemical potential, $\mu_B$ along the collision axis at RHIC. This effect, which is an outcome of the nuclear transparency effect [5] is incorporated by writing [27,28]

$$\mu_B = a + b\, y_0^2 \qquad (2)$$

Where $y_0$ ( = $\xi z$) is the rapidity of the expanding hadronic fluid element along the beam axis ( z-axis). The transverse velocity component of the hadronic fireball is assumed to vary with the transverse coordinate *r* in accordance with the Blast Wave model as [29]

$$\beta_T(r) = \beta_T^s \left(\frac{r}{R}\right)^n \tag{3}$$

Where $n$ is a velocity profile index and $\beta^s_T$ is the hadronic fluid *surface* transverse expansion velocity and is fixed in the model by using the following parameterization [5]

$$\beta_T^s = \beta_T^0 \sqrt{1 - \beta_z^2} \tag{4}$$

The transverse radius of fireball $R$ is parameterized as [5, 30]

$$R = r_0 \exp\left(\frac{-z^2}{\sigma^2}\right) \tag{5}$$

Where the parameter $r_0$ fixes the *transverse* size of the expanding hadronic matter at the freeze-out, along with $\sigma$ (width of the distribution) for different values of the longitudinal i.e. the $z$-coordinate [5]. The contributions of various heavier hadronic resonances [27, 31] which decay after the freeze-out has occurred, are also taken into account in our analysis. We also impose the criteria of exact strangeness conservation.

## 3. Rapidity Spectra

In Figure 1 we have shown the rapidity distributions of some strange particles produced at LHC energy of $\sqrt{s}_{NN}$ = 7.0 TeV. The model calculations are found to be in fairly good agreement with the experimental data. The available data is taken from CMS experiment at CERN LHC [32] and is shown by colored filled shapes while as our model calculations are shown by solid curves. The best fit is obtained by minimizing the distribution of $\chi^2 / dof$ given by [33],

$$\chi^2 = \sum_i \left(\frac{R_i^{\exp} - R_i^{theor}}{\varepsilon_i}\right)^2 \tag{6}$$

We have taken only statistical errors into consideration. The $\chi^2/dof$ for fitting the rapidity spectra are minimized with respect to the variables $a$ and $b$ whereas the values of T and $\beta^0_T$ are first obtained by fitting the corresponding $p_T$ - distributions. The $p_T$ distributions are not affected by the values of $a$ and $b$, instead these parameters have significant effect on the rapidity distribution shapes. The width of the matter distribution $\sigma$ has insignificant effect on the shape of the spectra and its value has been fixed as 5 [6]. The fit parameters obtained from the rapidity distributions of the three experimental data sets at $\sqrt{s}_{NN}$ = 7.0 TeV are given in Table 1 below.

| Particle | $a$(MeV) | $b$ (MeV) |
|---|---|---|
| $K_s^0$ | 1.25 | 7.13 |
| $\Lambda + \bar{\Lambda}$ | 0.90 | 6.45 |
| $\Xi + \bar{\Xi}$ | 0.70 | 5.52 |

**Table .1** Values of model parameters $a$ and $b$ obtained from fitting the rapidity distributions.

It is evident from Table 1 that the value of the baryonic chemical potential approaches to zero in these experiments in the rapidity range of 0 ± 2 units. At $\sqrt{s_{NN}} = 7.0$ TeV a smaller value of *a* and a larger value of *b* indicates a higher degree of nuclear transparency in accordance with equation 2. However, on the overall basis it can be said that these LHC experiments involving p-p collisions give a clear indication of the existence of a nearly baryon free matter owing to a high degree of nuclear transparency effect**.** Another evidence for this transparency also comes from [34] where the measured mid-rapidity anti-baryon to baryon ratio is found to be nearly equal to unity at various LHC energies. This fact is also supported by the nearly flat rapidity distributions in Figure 1.

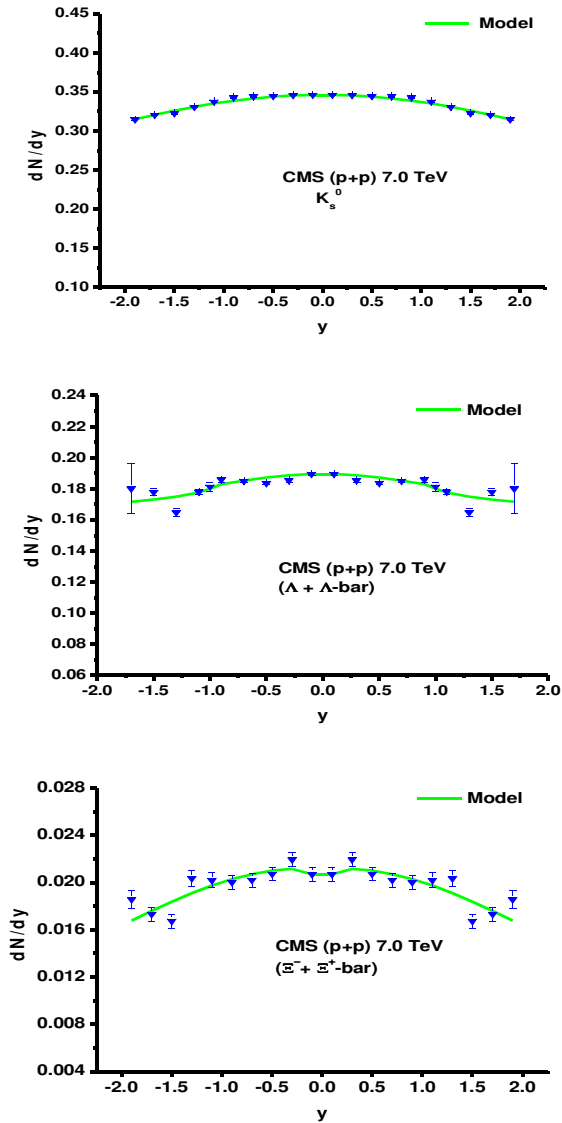

**Figure 1.** Rapidity distributions of strange hadrons produced at $\sqrt{s_{NN}} = 7.0$ TeV.

**4. Transverse momentum spectra**

In figure 2 the transverse momentum spectra of various hadrons produced in p-p collisions at $\sqrt{s_{NN}}$ = 7.0 TeV are shown. The solid curves represent our model calculations while as the experimental data points are represented by different colored shapes. The experimental data [35] is fitted quite well with our model calculations. The error bars here represent the sum of statistical and systematic uncertainties. The various freeze-out conditions obtained from the $p_T$ spectra of these hadrons are given in Table 2. From the Table, it is clear that the multi-strange heavier baryons freeze-out a little earlier than the singly strange and the other lighter non-strange particles. This phenomena, known as sequential freeze-out, is also seen in heavy ion collisions at RHIC and LHC [6,8]. However, it is not much apparent in case of p-p collisions as is evident from little differences between the freeze-out temperatures of lighter and heavier particles in Table 2. Also a significant collective flow is observed in p-p collisions which is found to decrease as one goes from lighter particles to heavier multi-strange particles. This fact may be attributed to their smaller interaction cross sections with the constituents of the medium due to which these particles freeze-out earlier and have a lesser time to develop collective effects as compared to lighter particles which have relatively greater cross sections. Also the value of the index parameter *n* is found to decrease towards heavier particles.

| Particle | $T$ (MeV) | $\beta^0_T$ | $n$ | $\chi^2 / dof$ |
|---|---|---|---|---|
| $K^+$ | 172 | 0.79 | 1.17 | 0.50 |
| $K^-$ | 173 | 0.79 | 1.17 | 0.53 |
| p | 174 | 0.83 | 1.10 | 1.85 |
| $\bar{p}$ | 175 | 0.83 | 1.10 | 2.10 |
| φ | 175 | 0.79 | 1.08 | 0.15 |
| $\Xi^-$ | 176 | 0.71 | 1.06 | 0.72 |
| $\Xi^+$ | 177 | 0.71 | 1.05 | 0.70 |
| Ω | 177 | 0.70 | 1.02 | 0.87 |
| $\bar{\Omega}$ | 178 | 0.70 | 1.02 | 0.47 |

**Table 2.** Freeze-out conditions of hadrons produced in p-p collisions at $\sqrt{s_{NN}}$ = 7.0 TeV.

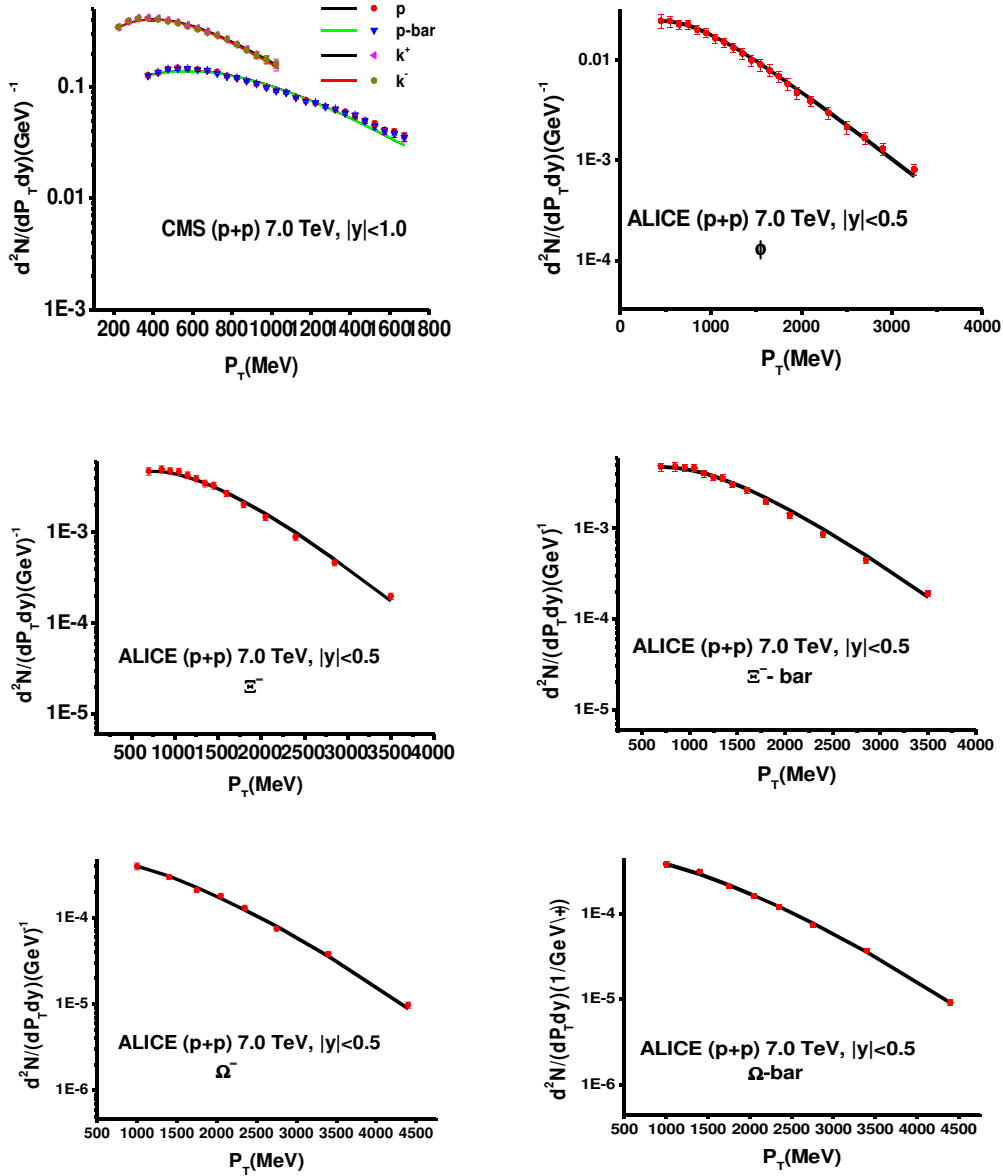

**Figure 2.** Transverse momentum spectra of various hadrons at $\sqrt{s_{NN}}$ = 7.0 TeV..

We have also successfully reproduced the transverse momentum distributions of protons and kaons at $\sqrt{s_{NN}}$ = 2.76 TeV, as shown in figure 3. The experimental data [35] is fitted quite well with our model calculations. The various freeze-out conditions obtained from the spectra are tabulated in Table 3. The error bars here represent the sum of statistical and systematic uncertainties. It is seen that the protons and kaons freeze-out simultaneously from the hadronic medium. Also a significant collective flow is observed in this case, but it is still lesser than that observed at $\sqrt{s_{NN}}$ = 7.0 TeV. This significant enhancement of collective flow with collision energy is understood to be due to the more particle production at higher energies.

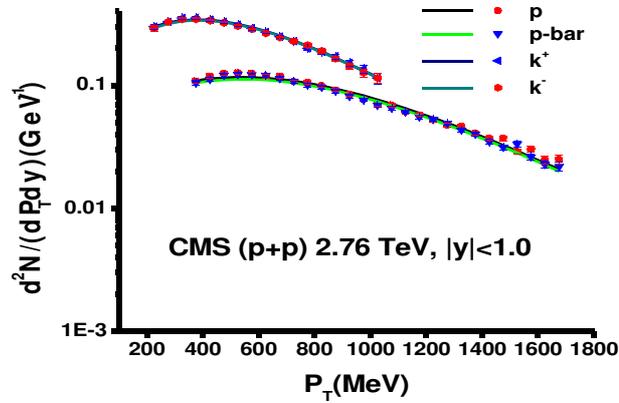

**Figure 3.** Transverse momentum spectra of protons and kaons at $\sqrt{s_{NN}}$ = 2.76 TeV.

| Particle | T(MeV) | $\beta^0_T$ | n | $\chi^2 / dof$ |
|---|---|---|---|---|
| p | 174 | 0.73 | 1.11 | 1.02 |
| $\bar{p}$ | 175 | 0.73 | 1.11 | 1.0 |
| $K^+$ | 173 | 0.70 | 1.18 | 0.41 |
| $K^-$ | 174 | 0.70 | 1.18 | 0.55 |

**Table 3.** Freeze-out conditions of protons and kaons $\sqrt{s_{NN}}$ = 2.76 TeV.

In order to compare the system properties in p-p collisions at LHC and RHIC, we have reproduced the transverse momentum distributions of p, $\bar{p}$, $K^+$ and $K^-$ produced in minimum bias p-p collisions at highest RHIC energy of $\sqrt{s_{NN}}$ = 200 GeV as shown in figure 4. The experimental data has been taken from STAR experiment [36]. A good agreement is seen between our model calculations (shown by solid curves) and the experimental data (shown by red shapes). The value of *n* comes out to be unity. The freeze-out conditions obtained by fitting the $p_T$ - spectra of these particles are given in table 4 below.

| Particle | $T$ (MeV) | $\beta^0_T$ | $\chi^2 / dof$ |
|---|---|---|---|
| p | 164.0 | 0.21 | 3.29 |
| $\bar{p}$ | 164.0 | 0.16 | 3.02 |
| $K^+$ | 165.0 | 0.15 | 1.89 |
| $K^-$ | 165.0 | 0.17 | 0.82 |

**Table 4.** Freeze-out parameters of protons and kaons $\sqrt{s_{NN}}$ = 200 GeV.

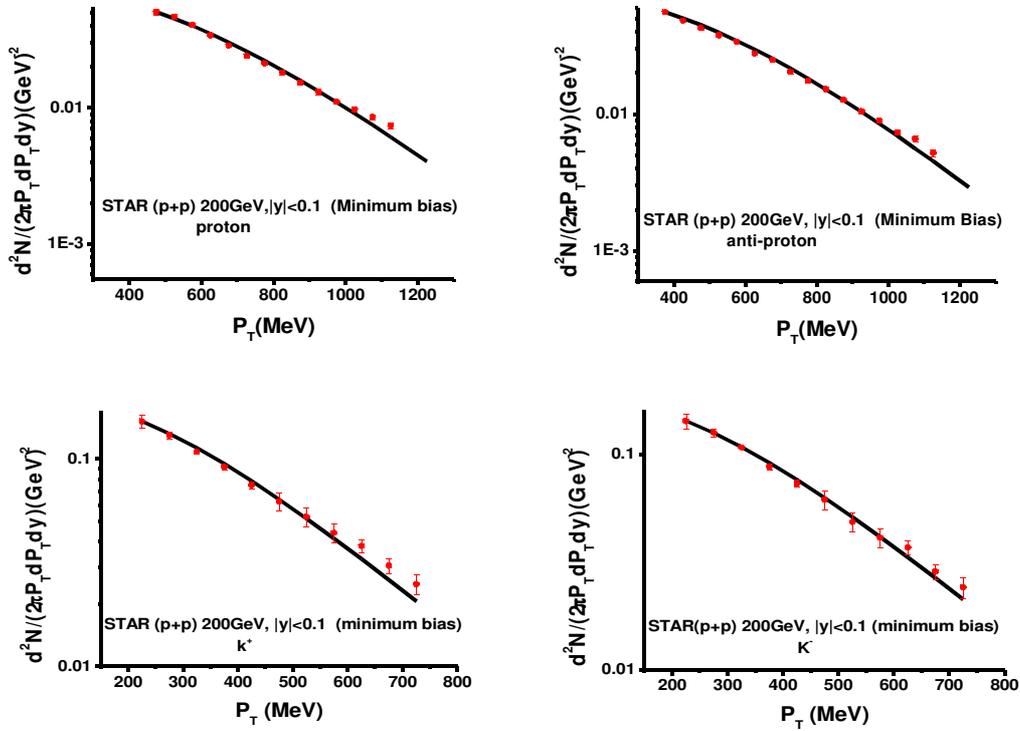

**Figure 4.** Transverse momentum distributions protons and kaons at $\sqrt{s_{NN}}$ = 200 GeV.

It is obvious from Table 4 that the hydrodynamical collective flow is almost insignificant in p-p collisions at RHIC. This is because of the less particle production which leads to a small system size such that the collective hydrodynamic effect at the final freeze-out is almost absent in the p-p collisions at the RHIC energies. On the other hand, the system formed in p-p collisions at LHC exhibits clear signature of a stronger hydrodynamical flow, as shown in Tables 2 and 3. Thus we can conclude that there is an onset of flow in the system formed in the p-p collisions at LHC which is because of the enhanced particle production, and hence a larger system size, taking place at these LHC energies compared to the highest RHIC energy. Moreover, in addition to the insignificant collective flow, the larger values of $\chi^2/dof$ obtained for protons and kaons at $\sqrt{s_{NN}}$ = 200 GeV indicates that the system is not in complete local thermal equilibrium as expected at LHC. Hence the formation of a hot and dense system attaining a reasonable degree of thermo-chemical equilibrium in a hadronic resonance gas phase before the final freeze-out takes place in p-p collisions at LHC. The system also possesses a significant collective hydrodynamic flow.

## 3. Summary and conclusion

The transverse momentum spectra of the various hadrons and the rapidity distribution of some strange hadrons produced in p-p collisions at $\sqrt{s_{NN}}$ = 7.0 TeV are fitted quite well by using our Unified Statistical Thermal Freeze-out Model (USTFM). The result extracted from the rapidity distributions of hadrons show that the chemical potential almost vanishes at $\sqrt{s_{NN}}$ = 7.0 TeV. This indicates the effects of almost complete transparency in p-p collisions at LHC. The LHC results show the existence of significant hydrodynamic flow present in the p-p system. A comparison between the transverse momentum spectra of protons and kaons at RHIC and LHC is made. For the system formed in p-p collisions at RHIC the collective behavior at final freeze-out is almost absent. We also observe a little earlier freeze- out of multi-strange particles as compared to lighter mass particles. This gives an indication of a sequential freeze-out , although less apparent, of various hadrons produced in the p-p collision system. Protons and Kaons are found to freeze-out almost simultaneously at all studied energies. The earlier freeze-out of heavier hyperons is indicated by their larger thermal temperature and smaller flow parameters. Also the transverse flow and freeze-out temperature are found to increase with collision collision energy from RHIC to LHC.

## Acknowledgements

The authors are gratefull to the University Grants Commision for providing financial assistance.